 \documentclass[final,authoryear,5p]{elsarticle}

\usepackage{epsfig}

\usepackage{amssymb}

\usepackage[ps2pdf,%
a4paper=true,%
breaklinks=true,%
colorlinks=true,%
pdfauthor={First Author et al.},%
pdftitle={Template for manuscripts in Advances in Space Research}%
]{hyperref}

\begin{document}

\begin{frontmatter}



\title{Multi-frequency study of large size radio galaxies 3C 35 and 3C 284}


\author[label1]{Dusmanta Patra\corref{cor}}
\ead{dusmanta.phy@gmail.com}
\author[label2]{Sabyasachi Pal}
\ead{sabya.pal@gmail.com}
\cortext[cor]{Corresponding author}
\address[label1]{S. N. Bose National Centre for Basic Sciences, Kolkata, 700106, India}
\address[label2]{Midnapore City College, Kuturiya, Bhadutala, 721129, India}


\begin{abstract}
\noindent{We report multi-frequency observations of two large radio galaxies, 3C 35 and 3C 284. The low-frequency observations were done with Giant Metrewave Radio Telescope starting from $\sim150$ MHz, and the high-frequency observations were done with the Very Large Array. We have studied the radio morphology of these two sources at different frequencies. We present the spectral ageing map using two of the most widely used models, the Kardashev-Pacholczyk and Jaffe-Perola models. Another more realistic and complex Tribble model is also used. We also calculate the jet power and the speed of the radio lobes of these galaxies. We check whether any episodic jet activity is present in these galaxies and found no sign of such activity.}
 
\end{abstract}


\end{frontmatter}

\parindent=0.5 cm
\section{Introduction}
\label{Sec:Intro}
{\noindent Radio galaxies (RGs) are astronomical objects with a lobe-jet structure hosted by optical galaxies situated between the radio lobes. From observations, it is noticed that each massive galaxy harbours super-massive black holes (SMBHs) \citep{Ri98}. 
A narrow collimated feature called `jets' connects the centre to the outer lobes, which are the indicators of the beams transporting energy from the core to the outer lobes. The jets are the bipolar relativistic outflows of ionised material originating from the vicinity of the SMBH. 
The overall linear sizes of RGs range from hundreds of kpc \citep{Br81, Al87, Ko06, Ma09} to over an Mpc scale \citep{Mu06, Da17,Da20,Oe22}, though giga-Hertz peaked spectrum sources and compact steep spectrum radio sources with sub-kpc scale structures are also found \citep{Go23,Od91}.
However, the linear size ($L$) of the powerful radio galaxies depends on their radio luminosity ($P$) and redshift ($z$). 
From the earlier work, examination of the angular size ($\theta$)-$z$ relationship suggested that $L$ decreases with $z$ \citep[e.g.][]{Mil71, Wa74}, but this relation could also be explained via an inverse correlation between $L$ and $P$ \citep{St77, Wa77}. Investigation of the {$\theta$}-$z$ relationship \citep[e.g.][]{Oo87,Si88} suggested that the physical sizes ($L$) of radio galaxies appear to depend on redshift by $L\propto(1+z)^{-3}$. 
The large size of these RGs means that they are not only capable of interacting with the interstellar medium but also with the environment in which the galaxy exists. These RGs are useful to understand the effects of electron energy loss in lobe plasma due to inverse-Compton scattering with Cosmic Microwave Background Radiation (CMBR) photons at various redshifts \citep{Ko04} and the spectral and dynamical ageing analyses to explain the evolution of sources \citep{Ko06, Ko08, Ja08, Ma09, Ma20}.

The beam model proposed by \cite{Lo73} is commonly accepted for the large-scale structure of these RGs. Later, \citet{Bl74} improved the beam model by implementing the ``twin-exhaust'' concept for double radio sources. According to this concept,  the active galactic nucleus is surrounded by too much dense thermal gas, preventing the relativistic plasma from escaping isotropically. The angular momentum defines an axis along which relativistic plasma generated in the nucleus flows along two oppositely-directed channels, and an equilibrium flow along this axis is established.
 Morphologically, there are two kinds of radio galaxies $-$ FR I and FR II \citep{Fr74}. The FR I type is of lower luminosity and has diffused, extended plumes of emission. This type of galaxy has reasonably symmetric large-scale jets and is the brightest closer to the core. FR II radio galaxies have higher luminosity compared to FR I and also have prominent hot-spots on the outer edges. Though recently \citet{Mi19} have studied a large sample of radio-loud active galactic nuclei (AGN) from the LOFAR Two-Metre Sky Survey, and they found that radio luminosity does not reliably predict whether a source is edge-brightened (FR II) or center-brightened (FR I).  The commonly accepted picture of FR II sources is that the relativistic jets interact with the surrounding medium to form a hot spot at the working surface. As the hot-spot progresses through the external medium \citep{Bu77, Bu79, Wi80, Me89}, the earlier accelerated plasma regions are compressed to form the typical lobes for FR II RGs \citep{Sc74, Be89, Ka97, Kr12}.

One of the key questions regarding large radio galaxies is whether their central activity is episodic or not. The central activity of the galaxies is related to the feeding of SMBHs, whose masses range from $\sim 10^6$ to $10^{10}M_{\odot}$. These SMBHs are present at the centres of most of the galaxies. 
It is not clear whether all galaxies sheltering SMBHs went through one or more AGN activities. Based on the current observations on SMBHs, \citet{Fr98} reported that multiple periods of such activities are possible. However, the time scales of these repeating activities are very large and are of the order of $10^9$ or $10^{10}$ year. Large angular size RGs are ideal for studying the radio source evolution and the probable episodic activity in these objects \citep{Sc00, Sa02}. Though episodic activity is not an issue for large radio galaxies alone and could be happening in compact sources as well \citep{Ba90,Sa09,Sh12,Kh16}.

We have selected two large angular-size radio galaxies, 3C 35 and 3C 284, for studying low-frequency radio morphology starting at 150 MHz using Giant Metrewave Radiowave Telescopes (GMRT). Archival Very Large Array (VLA) data is used to make higher frequency images of these two sources. Our object is to look for diffused emission at low frequencies and estimate the spectral ages of the lobes. We can estimate the most reliable injection spectral index ($\alpha_{\textrm{inj}}$) by combining the available low-frequency data with high-frequency data. We also check for any probable episodic activity present in these radio galaxies.

The galaxy 3C 35 is a giant FR II class radio galaxy and has a redshift of $z=0.0670$ \citep{Sp85}.
The projected linear size of 3C 35 is 869.2 kpc, which makes 3C 35 a giant radio galaxy \citep{Is99}.
The radio power of 3C 35 at 178 MHz is $P_{178}$ = $10^{24.99}$ W $Hz^{-1} sr^{-1}$ \citep{Ow89}.

The galaxy 3C 284 has a redshift of $z=0.2397$ \citep{Ad09}.
3C 284 is an FR II type large radio galaxy with a projected linear size of 674.4 kpc.
The radio luminosity of 3C 284 at 178 MHz is $P_{178} = 167 \times 10^{24}$ W Hz$^{-1}$ sr$^{-1}$ \footnote{https://3crr.extragalactic.info/cgi/sourcepage?103}.

The following cosmological parameters are used in the entire study: $H_0=67.8$ km s$^{-1}~\textrm{Mpc}^{-1}$, $\Omega_m=0.308$ and $\Omega_{vac}=0.692$ \citep{Ad16}.

\section{Observation, data reduction and spectral analysis}
\subsection{Observation by the GMRT}

The GMRT consists of 30 fully steerable antennas, each of which has a 45-meter diameter. Among the 30 antennas, 16 are located in a nearly `Y' shaped array, and the remaining 14 antennas are randomly distributed within about one-kilometre diameter to improve the sensitivity for extended emission. 
During the time of observations for the current paper, GMRT operated in five frequency bands centred at 151 MHz, 244 MHz, 325 MHz, 614 MHz, and from 1000 to 1450 MHz. GMRT observations were made at 153, 240, 325, and 606 MHz. 

The observation details, including frequency of observation, phase calibrator, observation date, and source observation time, are tabulated in Table \ref{obs-log}.
The observed bandwidth used was 32 MHz, except for the 153 MHz observation. 
Six MHz of bandwidth was used for the 153 MHz observation. 
The basic integration time used was 16.9 seconds.
\begin{table*}
\centering
\caption{Observing log}
\vskip 1cm
\label{obs-log}
\begin{tabular}{c c c c c c c}

\hline
Source    & Teles-     & Array  & Obs. & {    Phase }   & Obs.        & Obs. \\
          & cope       & Conf.  & Freq.& {    Calib.}   & Date        & Time \\
          &            &        & (MHz)  &              &             & (Min) \\
(1)     &  (2)       & (3)    &  (4) & {    (5)  }    & (6)         &    (7)  \\
\hline
3C 35      & GMRT       &        & ~153  & {     3C 48}    & 2007 Dec 07 & 179\\
3C 35      & GMRT       &        & ~240  & {     3C 48}    & 2007 Nov 03 & 407\\
3C 35      & GMRT       &        & ~325  & {     3C 48}    & 2008 Feb 23 & 296\\
3C 35      & GMRT       &        & ~606  & {     3C 48}    & 2007 Nov 03 & 407\\
3C 35      & VLA        & C      & 1422 & {    0111+481} & 1997 Aug 27 & 168 \\
3C 284     &  GMRT      &        & ~153  & {    3C 286   } & 2007 Dec 08 & 278\\
3C 284     &  GMRT      &        & ~240  & {    3C 286   } & 2008 Mar 07 & 344\\
3C 284     &  GMRT      &        & ~325  & {    3C 286   } & 2008 Feb 23 & 333\\
3C 284     &  GMRT      &        & ~614  & {    3C 286   } & 2008 Mar 07 & 344\\
3C 284     &  VLA       &  B     & 1527 & {    1328+254} & 1985 May 12 & ~30\\
3C 284     &  VLA       &  D     & 8265 & {    1308+326} & 1992 Aug 22 & ~21\\
\hline
\end{tabular}
\end{table*}

The Astronomical Image Processing System ({\tt AIPS})\footnote{http://www.aips.nrao.edu} was used for the reduction of the data. \cite{Pe13} flux density scale was used.
Perley \& Butler scale is based on the absolute flux density measurements determined by combining flux density ratios between a group of potential calibrators and the planet Mars by the VLA. According to observations, the radio sources 3C 123, 3C 196, 3C 286, and 3C 295 exhibit minimal variation, at a level of less than approximately 5\% over a century, across a wide range of frequencies from 1 to 50 GHz. This makes them excellent candidates as reference standards for measuring the flux density of other radio sources \citep{Pe13}.

Time and channel-based flagging for data affected by radio frequency interference were performed using a 6$\sigma$ threshold median filter. Based on closure errors on the bandpass calibrator, a few baselines were removed. For certain time ranges, the data for antennas with high errors in antenna-based solutions was examined and removed. After a few rounds of imaging and self-calibration, residual errors above 5$\sigma$ were also flagged. The system temperature ($T_{sys}$) was discovered to vary with antenna and elevation. In the absence of regular $T_{sys}$ measurements for GMRT antennas, this correction was computed from the residuals of corrected data with respect to the model data. After that, the corrections were applied to the data.
The final image was made after several rounds of phase-based self-calibration, and one round of amplitude-based self-calibration, where the data was normalised by the median gain for all the data.

\subsection{Observation by the VLA}

For higher frequency observations, available archival VLA data was used. The VLA consists of twenty-seven antennas arranged in a Y-shaped array, each with a diameter of 25 metres. For 3C 35, the VLA C configuration is used for 1422 MHz observation. For 3C 284, VLA B and D configurations are used for 1527 and 8265 MHz observations, respectively. 

For data analysis, {\tt AIPS} is used. For the analysis of 3C 35 data, 3C 48 is used as the flux-density calibrator and for the analysis of 3C 284 data, 3C 286 is used as the flux-density calibrator. \cite{Pe13} flux density scale is used. Both of the sources were amplitude-based self-calibrated a few times and phase based self-calibrated once. We checked that there is no sign of a systematic shift in the positions of the point sources before and after self-calibration.

\subsection{Spectral analysis}
For spectral analysis, we used the Broadband Radio Analysis ToolS (BRATS) software package \citep{Ha13, Ha15}. The BRATS software is developed using the C programming language, and the PGPLOT graphics subroutine library is used for visualization. The radio maps are aligned in the same geometry at each frequency, and the images are convolved with the same resolution. The same resolution radio maps are loaded into BRATS with a 5 sigma cutoff depending upon the RMS value. We exclude the core of the galaxy for spectral analysis as it is not a region of interest. For strong sources, the error in flux measurement is dominated by systematics in calibration. This can be calculated from the quadrature sum of the systematic, noise and fitting error. For GMRT, we have taken a conservative 10\% of systematic errors, while for VLA, 2\% of systematic errors. To find the best fitting injection index for a minimum reduced chi-square value, {\tt findinject} command of BRATS is used. A best-fitting injection index of 0.50 for 3C 35 and 0.90 for 3C 284 is found. The final model fitting of the sources was done in BRATS using the obtained values of the injection index. 
For 3C 35, \citet{Or10} have reported an average injection spectral index of 0.50, which is consistent with our estimated value.
However, \citet{La80} calculated steeper spectra for 3C 35 with a value of 0.89, which may have appeared because their data sensitivity was insufficient. For the galaxy 3C 284, \citet{La80} determined a spectral index of $0.82$, which is slightly flatter compared to our calculated value of 0.90.

\begin{figure*}
\begin{center}
\includegraphics[width=15 cm]{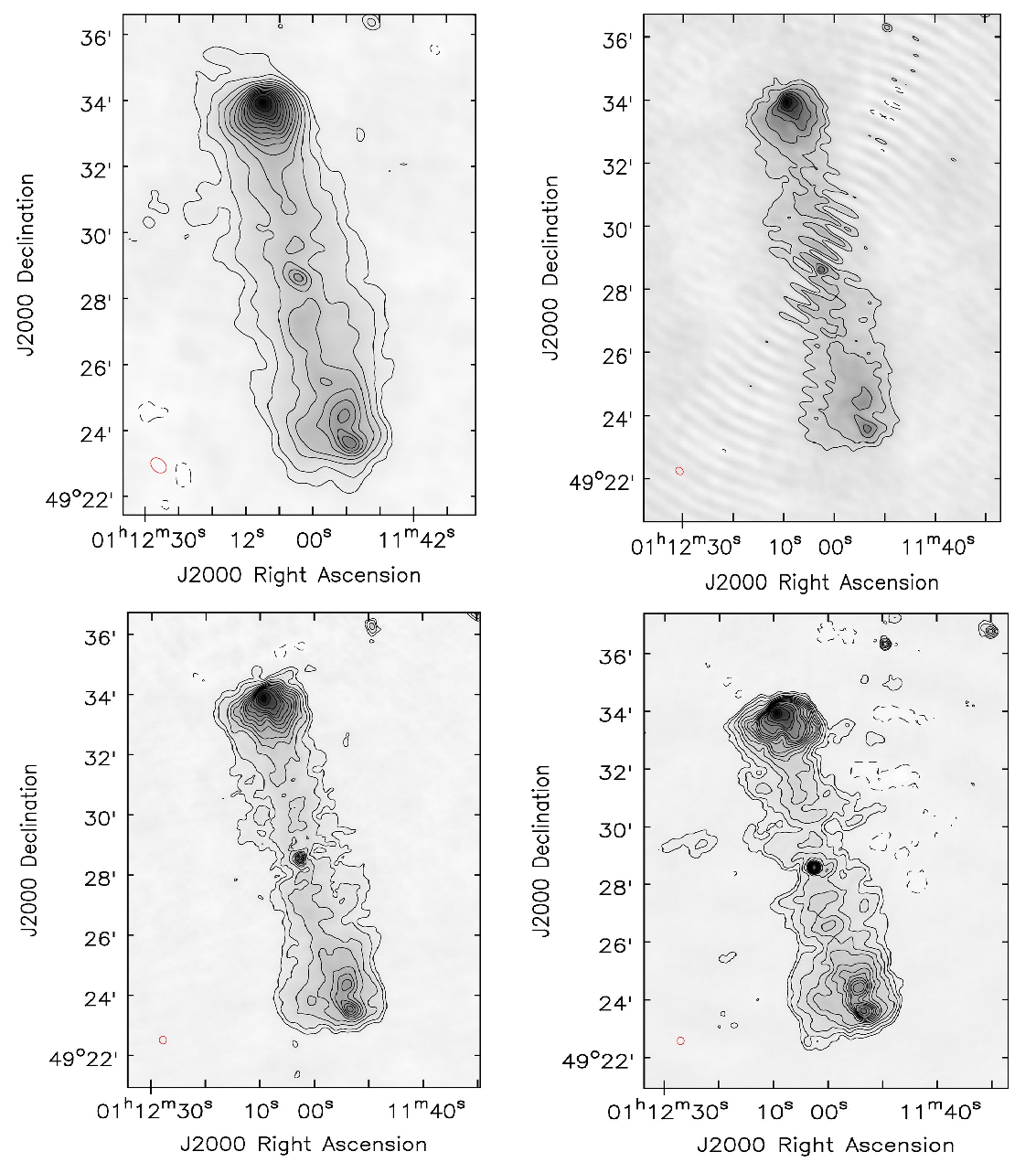}
\caption{GMRT low-frequency images of 3C 35 at 153, 240, 325 and 606 MHz. The contours are drawn at $3 \sigma$ $\times$ (--1, 1, 2, 4, 6, 8, 10, 12, 14, 16, 18 ...), where $\sigma$ is the local RMS noise, tabulated in Table \ref{Obs-result-galaxy}. The negative contour is shown with the dashed line. The restoring beam is shown by an ellipse in the lower left corner of each image in red colour.} 
\label{3C35}
\end{center}
\end{figure*}

\section{Results and Discussion}

\begin{table*}
\centering
\scriptsize
\caption{The observational parameters and observed properties of the sources}
\vskip 0.2cm
\begin{tabular}{c c cc c c c c}

\hline
Source & Freq. & \multicolumn{2}{c}{Beam } & Position Angle      & RMS        & $S_P$             & $S_I$    \\
       & (MHz) & ($^{\prime\prime}$)       & ($^{\prime\prime}$) & ($^\circ$) & (mJy beam$^{-1}$) & (mJy beam$^{-1}$)& (mJy)        \\
\hline
3C 35  & ~153  & 31.4  & 22.9 & ~45   & ~6.3  & ~670  & 20542    \\
       & ~240  & 15.8  & 12.2 & ~51   & ~4.1  & ~160  & 12990     \\
       & ~325  & 10.7  &  9.5 & ~72   & ~1.1  & ~~91  & 12676     \\
       & ~606  & ~5.2  &  3.8 & ~48   & ~0.5  & ~~86  & ~5275     \\
       & 1422  & 14.3  & 13.3 & --86  & ~0.3  & ~~39  & ~2308     \\

3C 284 & ~153  & 28.2  & 25.8 & 0.8  &  23.2  & 5320 & 20472    \\
       & ~240  & 18.2  & 16.9 & 1.8  &  16.3  & 2423 & 12031    \\
       & ~325  & 12.5  &  9.9 & ~76  &  14.1  & 2169 & 11920    \\
       & ~614  & ~5.2  &  4.5 & ~55  &  ~2.1  & ~566 & ~4934   \\
       & 1527  & ~4.9  &  4.1 & ~83  &  ~0.2  & ~226 & ~1915   \\
       & 8265  & 16.3  &  7.5 & ~85  &  ~0.8  & ~~81 & ~~324    \\
\hline       
\end{tabular}
\label{Obs-result-galaxy}
\end{table*}

The GMRT images of 3C 35 are shown in Figure \ref{3C35}, while those of 3C 284 are shown in Figure \ref{3C284}. Table \ref{Obs-result-galaxy} summarises the properties of images of these two sources at different frequencies. The name of the source (column 1), observational frequency in MHz (column 2), the major and minor axes of the restoring beam in arcsec (columns 3--4), position angle (PA) of beams in degrees (columns 5), the RMS noise in mJy beam$^{-1}$ (columns 6) and the estimated peak and integrated flux densities at different frequencies (columns 7--8) are shown in the Table \ref{Obs-result-galaxy}. Observations at frequencies greater than 1 GHz are done using VLA, and the rest of the observations are done with GMRT.

\begin{figure*}
\begin{center}
\includegraphics[width=17 cm]{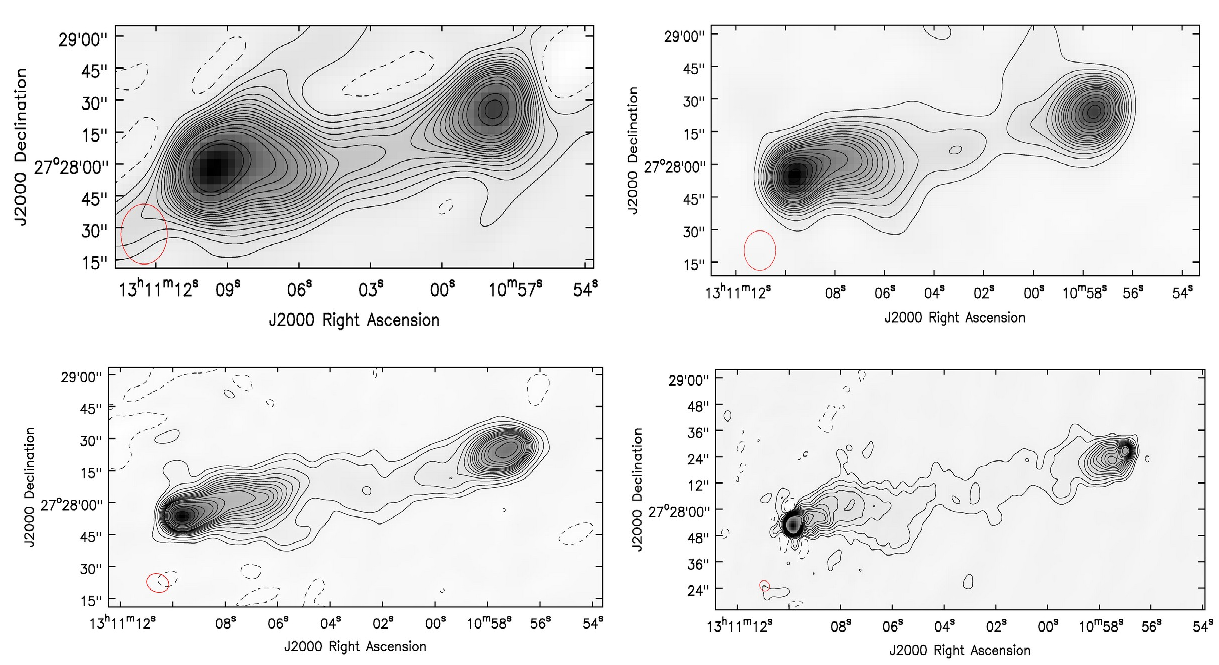}
\caption{GMRT low-frequency images of 3C 284 at 153, 240, 325 and 614 MHz. The contours are drawn at $3 \sigma$ $\times$ (--1, 1, 2, 4, 6, 8, 10, 12, 14, 16, 18...), where $\sigma$ is the local RMS noise, tabulated in Table \ref{Obs-result-galaxy}. The negative contour is shown with the dashed line. The restoring beam is shown by an ellipse in the lower left corner of each image in red colour.} 
\label{3C284}
\end{center}
\end{figure*}

\subsection{Radio morphology of the source}
3C 35 is a classic example of a large-sized FR II type radio galaxy extended along the north-south direction.
Previously the source was studied at frequencies 608 MHz, 1.4 and 5 GHz using the Westerbork Synthesis Radio Telescope
\citep{va82, Ja87, Sc00b} and 74 and 327 MHz with the VLA \citep{Or10}.

From Figure \ref{3C35}, it is noticed that 3C 35 has a similar type of morphology along both sides of the galaxy core. The distance between the northern and southern lobes from the core of the galaxy is nearly the same. The radio emission of the southern lobe is less bright in comparison to the northern lobe. \citet{va82} analyzed images of the radio source at 5 GHz and suggested that the south hot spot in the source could be a double. However, upon examining our 606 MHz GMRT image, the second weaker hot spot is more likely a knot in the lobe rather than a different hot spot. The source was observed at 327 MHz with VLA B configuration by \citet{Or10}. Considering their image at 327 MHz, \citet{Or10} also noted that the second weak hot spot may be a knot.
\citet{Or10} found a total flux density of 11.7 Jy which is a little bit low compared to our measured total flux density of 12.6 Jy at 325 MHz. This may be due to the missing flux at the VLA B configuration in \citet{Or10}.

For galaxy 3C 284, the lobes are elongated in the east-west direction. The length of the eastern and western lobes of the galaxy is nearly similar, making both of the lobes symmetric. From Figure \ref{3C284}, it is observed that the eastern lobe is much brighter than the western lobe.

The reasons behind the asymmetric morphology in the lobes of galaxies are linked with the environment of the galaxy.
If the gas density of the external medium through which the jet propagates is different at both sides of the core of the galaxy, the lobe morphologies may differ.
If the jet goes through a low-density environment, the jet easily propagates, and the structure becomes elongated. The flux density difference in both of the jets may also be due to Doppler boosting.

\subsection{Magnetic field of the sources}
We calculated the magnetic fields of the sources using the formalism suggested by \cite{Ko08}.
Assuming the minimum energy condition, the magnetic field in micro Gauss is given by the following equation
\begin{equation}
\label{mag-eq}
B_{min}=10^{6}(1+z) \bigg[\frac{3.26604 \times 10^{-23}}{\sin^{3/2}\phi} \frac{(1+k)A^\prime}{\eta f_s \theta^{\prime\prime}_x \theta^{\prime\prime}_y s}\bigg]^{2/7}
\end{equation}
where $A^\prime=C_4\int^{\nu_{2}}_{\nu_{1}} \frac{S (\nu)}{\sqrt{\nu}}d\nu$.
$C_4$ is a constant with a value of $1.05709 \times 10^{12}$ in cgs units defined by \cite{Pa70}.
Here, the symbols $k$ is the proton to electron energy density ratio, $\eta$ is the filling factor, $f_s$ is a shape factor for the volume of the emission region, $\theta^{\prime\prime}_x$ and $\theta^{\prime\prime}_y$ are the two projected dimensions of the emission region, and $s$ is the depth of the emission region in kpc.
It is assumed that the volume of the emission region has a cylindrical shape factor, and the value of $f_s$ is taken to be $\frac{\pi}{4}$.
$S(\nu)$ is the flux density in mJy at frequency $\nu$ (MHz). For the integration limits, a range of 10 MHz to 100 GHz is used in the frame of the emitter.
$(2/3)^{3/4}$ is substituted for $\sin^{3/2}\phi$ in the equation \ref{mag-eq} to calculate $B_{min}$.
Using the above equation, the calculated magnetic fields for 3C 35 and 3C 284 are $1.1 \times 10^{-9}$ Tesla and $3.0 \times 10^{-9}$ Tesla respectively.

The choice of the geometry of the lobes of radio galaxies may affect the accuracy of calculating magnetic fields. In this paper, we assumed a cylindrical model for the lobes, a common simplification used in radio astronomy. While this model is relatively simple to use and provides reasonable estimates of the magnetic fields, it may not accurately represent the true geometry of the lobes. A conical geometry may be more realistic for some radio galaxies, particularly those with more complex morphologies. However, using a conical model may require additional assumptions about the opening angle and axis of the cone, which could introduce additional uncertainties in the magnetic field calculation. In general, the choice of geometry should be based on the specific characteristics of the radio galaxy being studied. More complex geometries may be necessary for radio galaxies with complex morphologies, while simpler geometries may be sufficient for simpler sources. The two sources, 3C 35 and 3C 284, have a simple morphology, and a cylindrical model is good enough for them.
\begin{figure*}
\label{specage-3C35}
\centering
\includegraphics[angle=0,width=8.8cm]{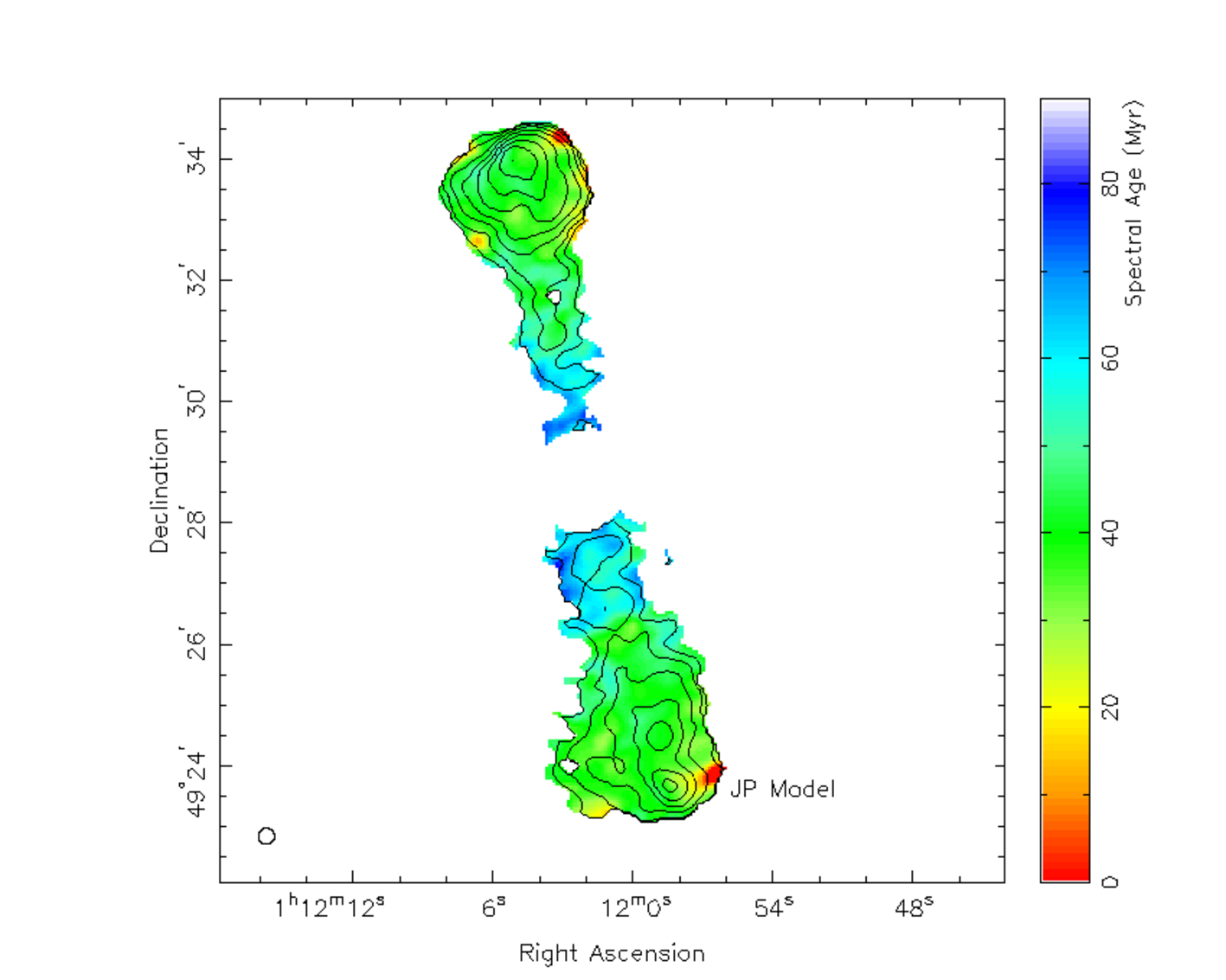}
\includegraphics[angle=0,width=8.8cm]{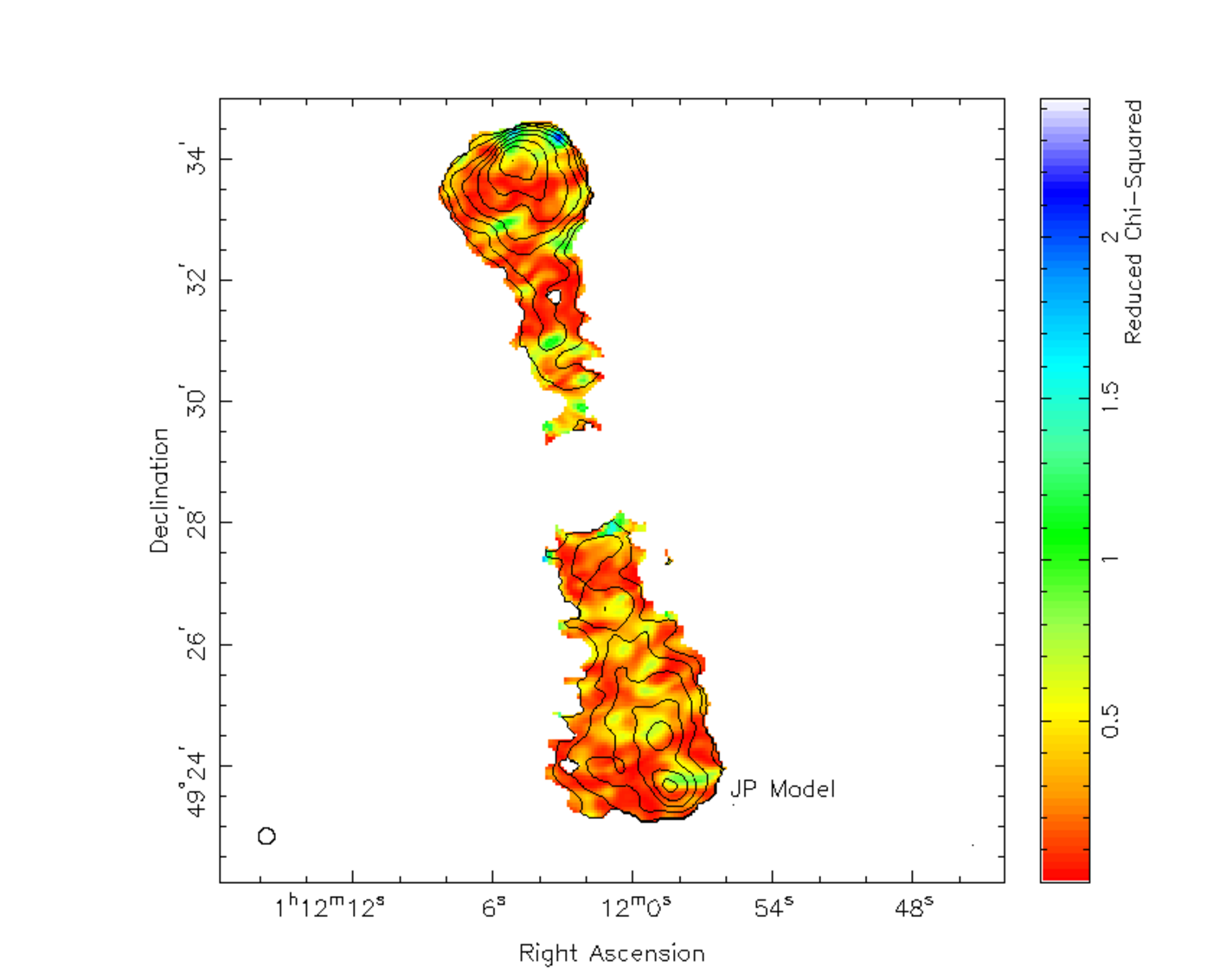}
\includegraphics[angle=0,width=8.8cm]{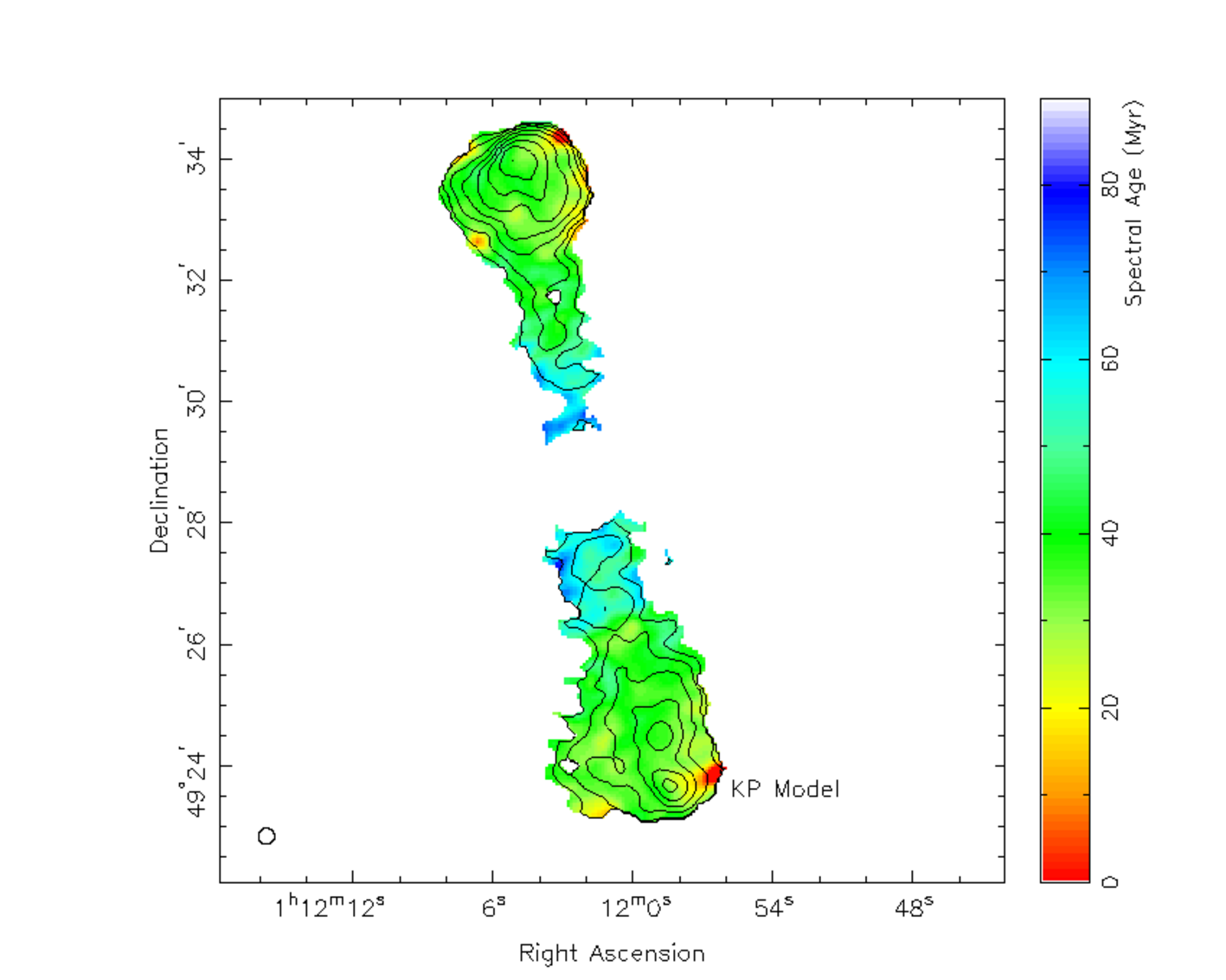}
\includegraphics[angle=0,width=8.8cm]{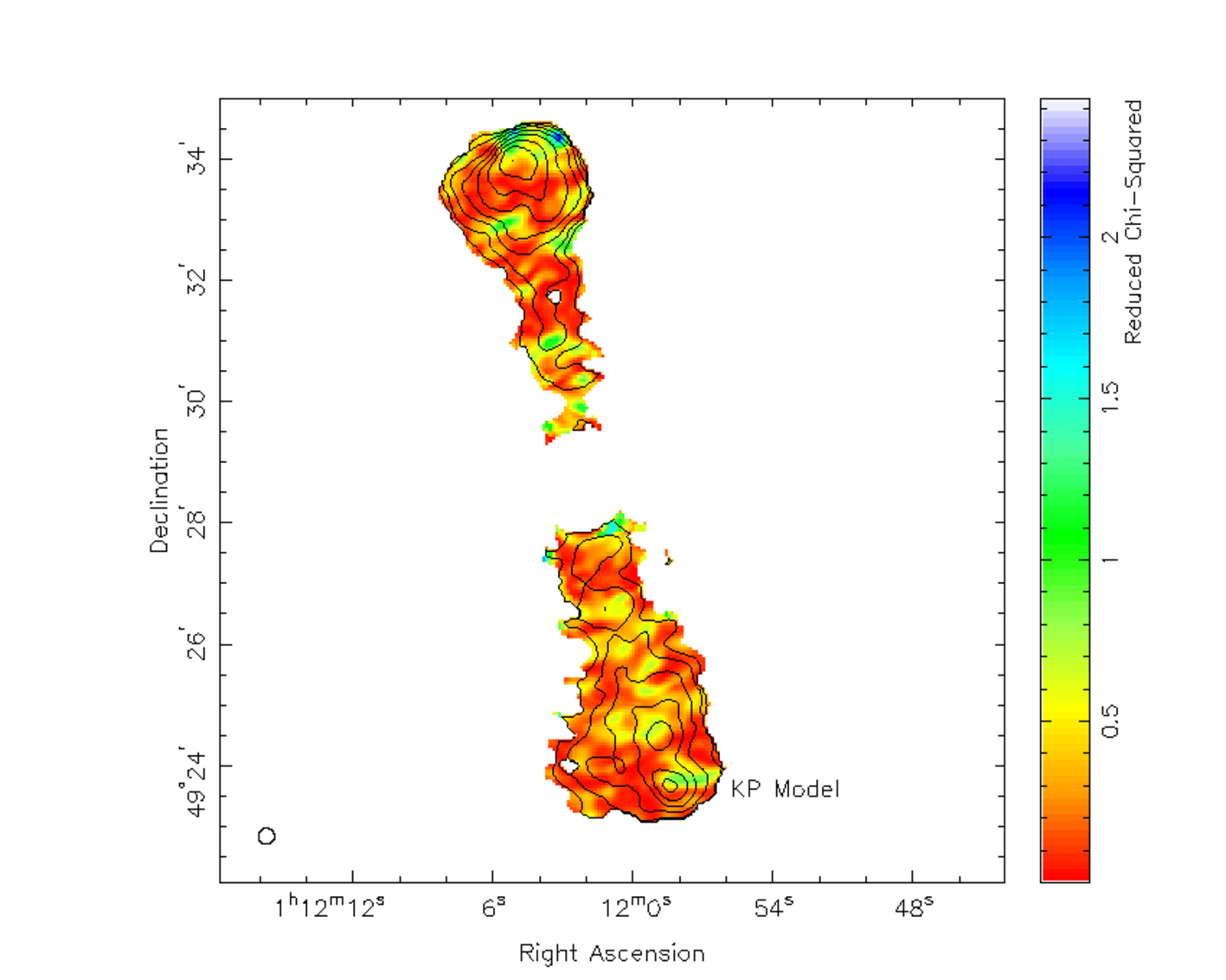}
\includegraphics[angle=0,width=8.8cm]{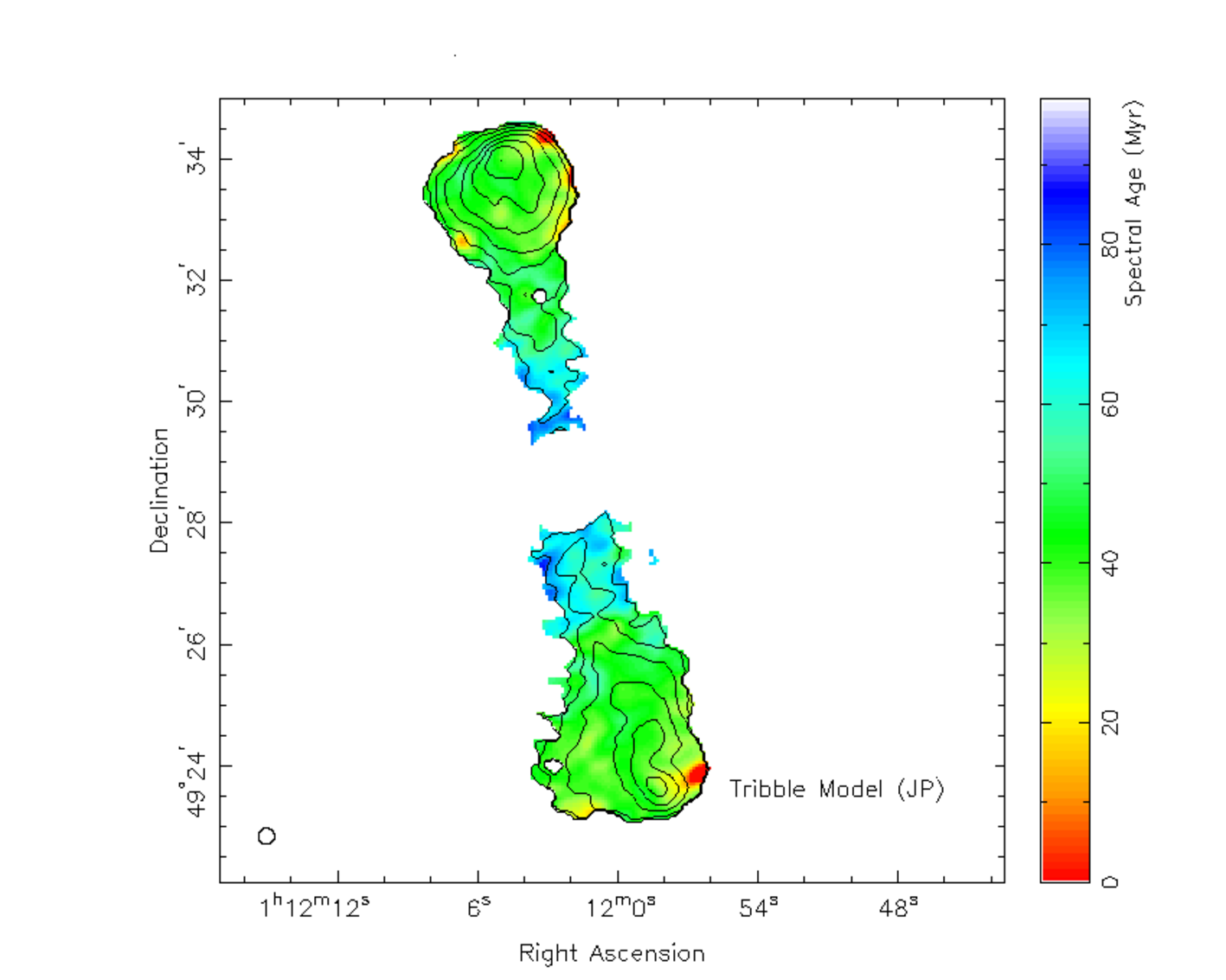}
\includegraphics[angle=0,width=8.8cm]{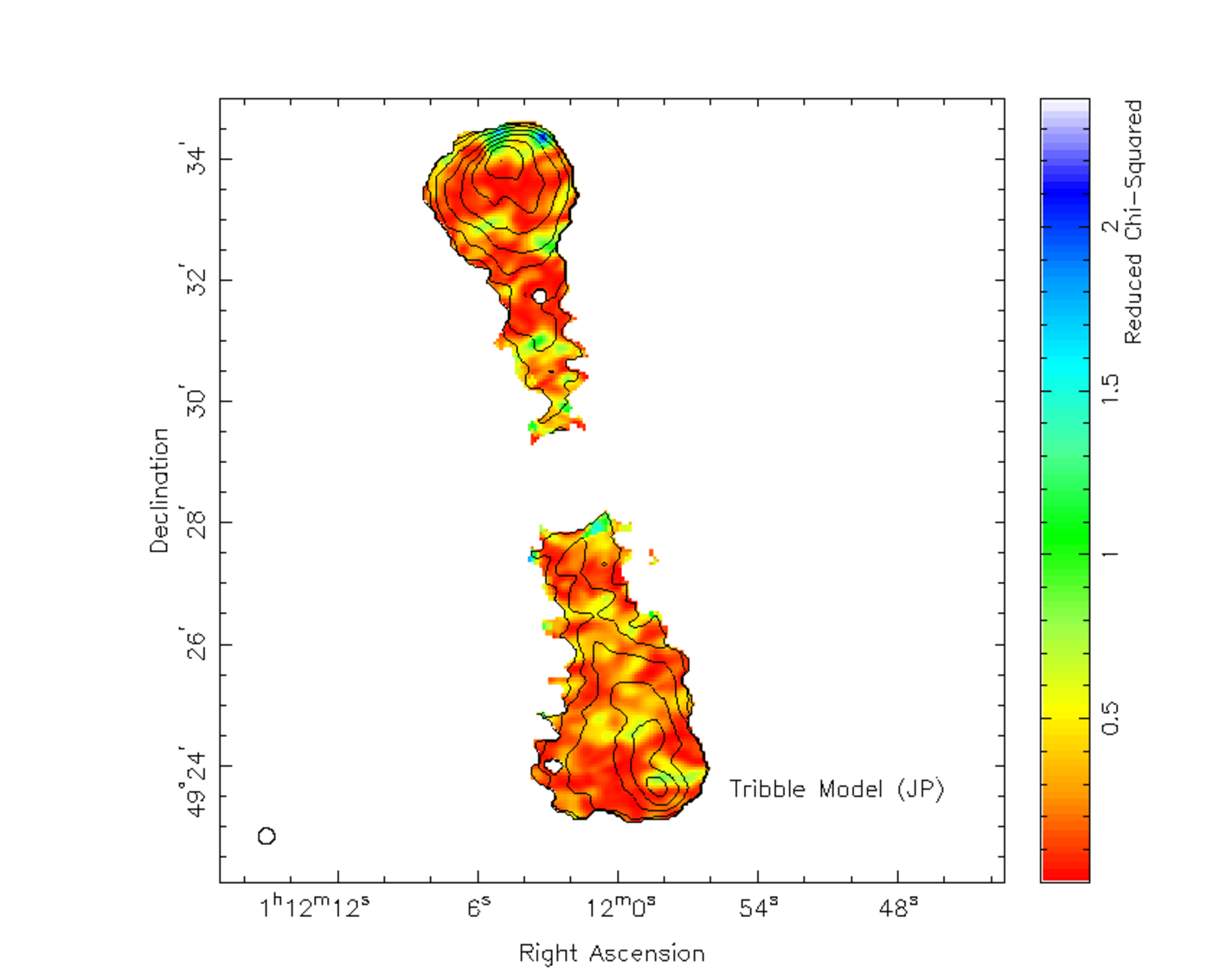}
\caption{Spectral ageing maps (left) and corresponding $\chi^2$ maps (right) of 3C 35 with 1.4 GHz flux contours. Three model fits are shown for the KP model (top), the JP model (middle) and the Tribble model (bottom) with an injection index of 0.50.}
\label{spec-age-3C35}
\end{figure*}

\subsection{Spectral ageing analysis}
Synchrotron emission from a region with a fixed magnetic field losses energy in the form of $\frac {dE}{dt} \propto \nu^2$. So, the cooling of higher energy electrons results in a curved and steeper spectrum in the plasma of the older region. Spectral ageing has become a widely utilised technique to understand the  mechanisms involved in the radio jets of different galaxies. The earlier work on the spectral ageing model was discussed regarding the physical reality of this model \citep[e.g.][]{Al87, Ma09, Ma10, Na10} to measure the age and hence dynamics of the powerful radio galaxies.

The models suggested by \cite{Ka62} and \cite{Pa70} (the KP model) and by \cite{Ja73} (the JP model) are widely used for spectral ageing analysis. Both of the above models are assumed to have a single injection of electron energy distribution, considered to be generated at the point of termination of the jet (e.g., \citet{My85}; \citet{Me89}; \citet{Ca91}). In the KP model, the assumption is that the pitch angle is constant for each electron but different for different electrons, while in the JP model, it is assumed that the pitch angles are to be scattered and isotropic on a time-scale that is much smaller in comparison with the spectral age of the source.

The initial distribution of the electrons is given by the power law.

\begin{equation}
~~~~~~~~~~~~~~~~~~~~~~N(E)=N_0 E^{-\delta}
\end {equation}
where $N_0$ and $\delta$ are normalisation constant and power law index, respectively, for the initially injected electron energy distribution. The backflow is, therefore, due to synchrotron losses as it moves around the lobes, eventually allowing the age and hence the power of these outflows to be determined. Because of synchrotron and inverse-Compton losses, the electron energy distribution $N$, for the KP model is given by \cite{Pa70} 
\begin{equation}
~~~~~~~~~N(E,\theta,t)=N_0E^{-\delta}(1-E_TE)^{\delta-2}.
\end{equation}
The intensity at a given frequency is given by
\begin{equation} 
\begin{array}{cc}
I_{\nu}(t) = 4\pi C_{3} N_{0} sB \int^{\pi / 2}_{0}d\theta \sin^{2} \theta \int^{E_{T}^{-1}}_{0} dE F(x) E^{-\delta} \\~~~~~~~~~~~~\times (1 - E_{T} E)^{\delta -2},
\end{array}
\end{equation}
where 
\begin{equation}
\label{ET-KP}
E_{T} \equiv C_{2} B^{2} (\sin^{2} \theta) t.
\end{equation} 
Here $E$ is the energy, $\theta$ is the pitch angle, and $t$ is the time since the electron was last accelerated. $C_2$, $C_3$, $\nu_c$ and the function $F(x)$ are explained by \cite{Pa70}. From the equation \ref{ET-KP}, it is noticed that the pitch angle $\theta$ is isotropic and constant throughout the radiative lifetime of the electrons. However, for the JP model \cite{Ja73}, it is assumed that the pitch angle is isotropic only at short time scales compared to the radiative lifetime, thus giving 
\begin{equation}
E_{T} \equiv C_{2} B^{2} \langle \sin^{2}  \theta \rangle t
\end{equation} 
where  $\langle \sin^{2} \theta \rangle$ stands for the time averaged pitch angle.

One of the primary assumptions of the JP and KP models is that the magnetic field is constant throughout the source. But this is unlikely to be a real case. A model was proposed by Tribble in which the magnetic field is varied throughout the source \citep{Tr93}. This model is more realistic, allowing electrons to diffuse across the layer at varying magnetic field strengths. 
\citet{Tr93} studied the spectral ageing of radio synchrotron emission from relativistic electrons in a random magnetic field. He considered a range of break frequencies corresponding to the range in field strength.
\citet{Ha13} has used the Tribble model for performing spectral ageing analysis for FR II radio galaxies. They conclude that the Tribble model gives both an excellent fit to observations and a physically realistic picture of the source.

We have used the BRATS software package for spectral ageing analysis of the radio galaxies 3C 35 and 3C 284. We present the spectral ageing maps of these sources in Figure \ref{spec-age-3C35} and \ref{spec-age-3C284}, respectively. The findings of these two sources are discussed in the following two sub-sections.   

\subsubsection{3C 35}
The spectral age and $\chi^2$ values as a function of position for the JP, KP, and Tribble models are shown in Figure \ref{spec-age-3C35} for the radio galaxy 3C 35. 
It is noticed that the spectral age varies throughout the lobes. 
The spectral age of hot-spots is less than the rest of the lobe, and it is increasing along with the jets towards the core. 
It is noticed that the spectral morphology is like a classical case, with the lowest age regions located close to the hot-spots and a usual trend of increasing while moving towards the core of the galaxy.
The age of the hot-spots is close to 40 Myr, while the age of the other parts of the lobe varies around 60 Myr. 
Across the width of the lobes, we noticed that the age variation is non-negligible. 
A discrete low age region is observed in the southern lobe of the galaxy. 
These low age regions in the hot-spots are because of a relic of an old hot spot or fast outflow.
We also noticed some of the regions at the top edge of the northern lobes with relatively high $\chi^2$. 
The relatively poor model fit in these regions is due to edge effects and dynamic range issues.

Upon comparing our findings with the results from \citet{Or10}, we have identified differences in the calculated synchrotron ages for the North and South lobes of 3C 35 by \citet{Or10} and measurements reported in the present paper. In their study, \citet{Or10} reported higher age values of $138^{+20}_{-22}$ million years for the North lobe and $147^{+35}_{-36}$ Myr for the South lobe, which differs from our estimates. The possible reason behind these discrepancies lies in the data quality used in their study. \citet{Or10}   calculated the spectral age of 3C 35 using four frequency data points (74, 327, 608, and 1400 MHz) but at a relatively poor resolution of 95 arc seconds. In contrast, in our study, we have used a much better resolution of 16 arc seconds, which should lead to more accurate results. Additionally, \citet{Or10} combined data at 327 MHz from the VLA in the B configuration with low-resolution data from the VLA in the C configuration $(55'' \times 50'')$. This combination might have caused missing flux issues in their data due to the VLA B and C configurations. In contrast, our data from the GMRT was collected using a hybrid configuration, significantly reducing the chances of missing flux problems.
Taking into account these factors, the differences in age measurements between \citet{Or10} and our study may have been raised.
\begin{figure*}
\centering
\includegraphics[angle=0,width=8.8cm]{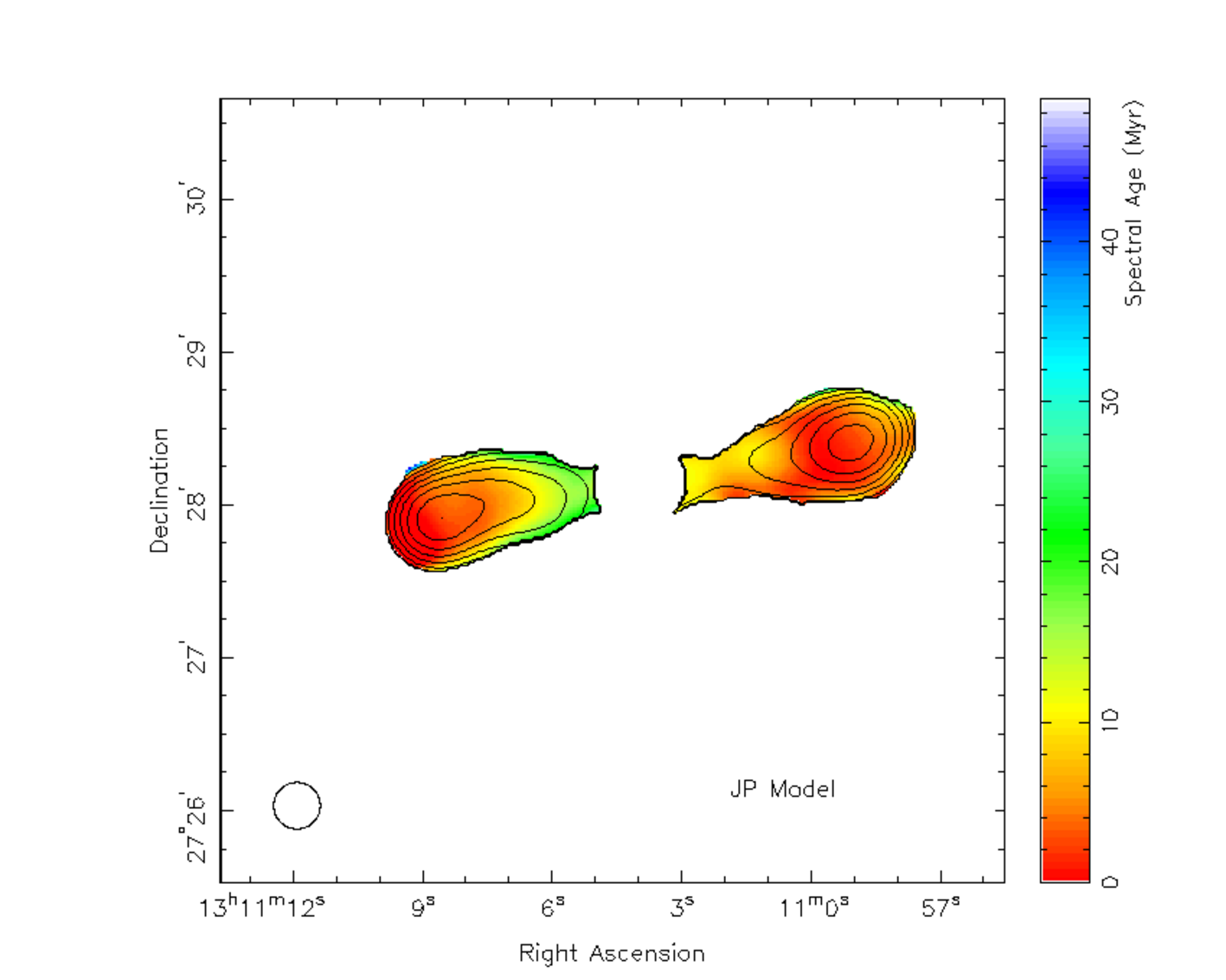}
\includegraphics[angle=0,width=8.8cm]{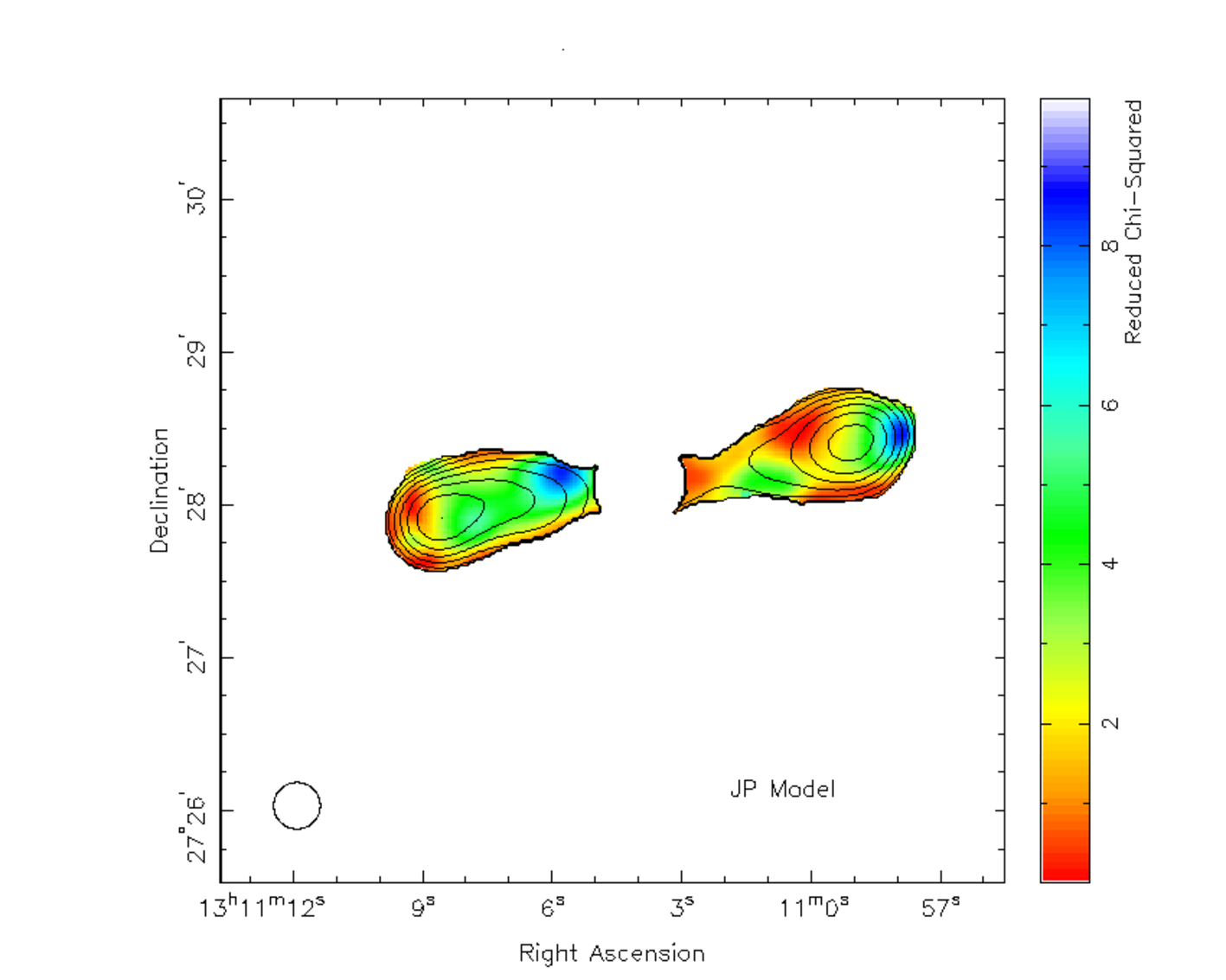}
\includegraphics[angle=0,width=8.8cm]{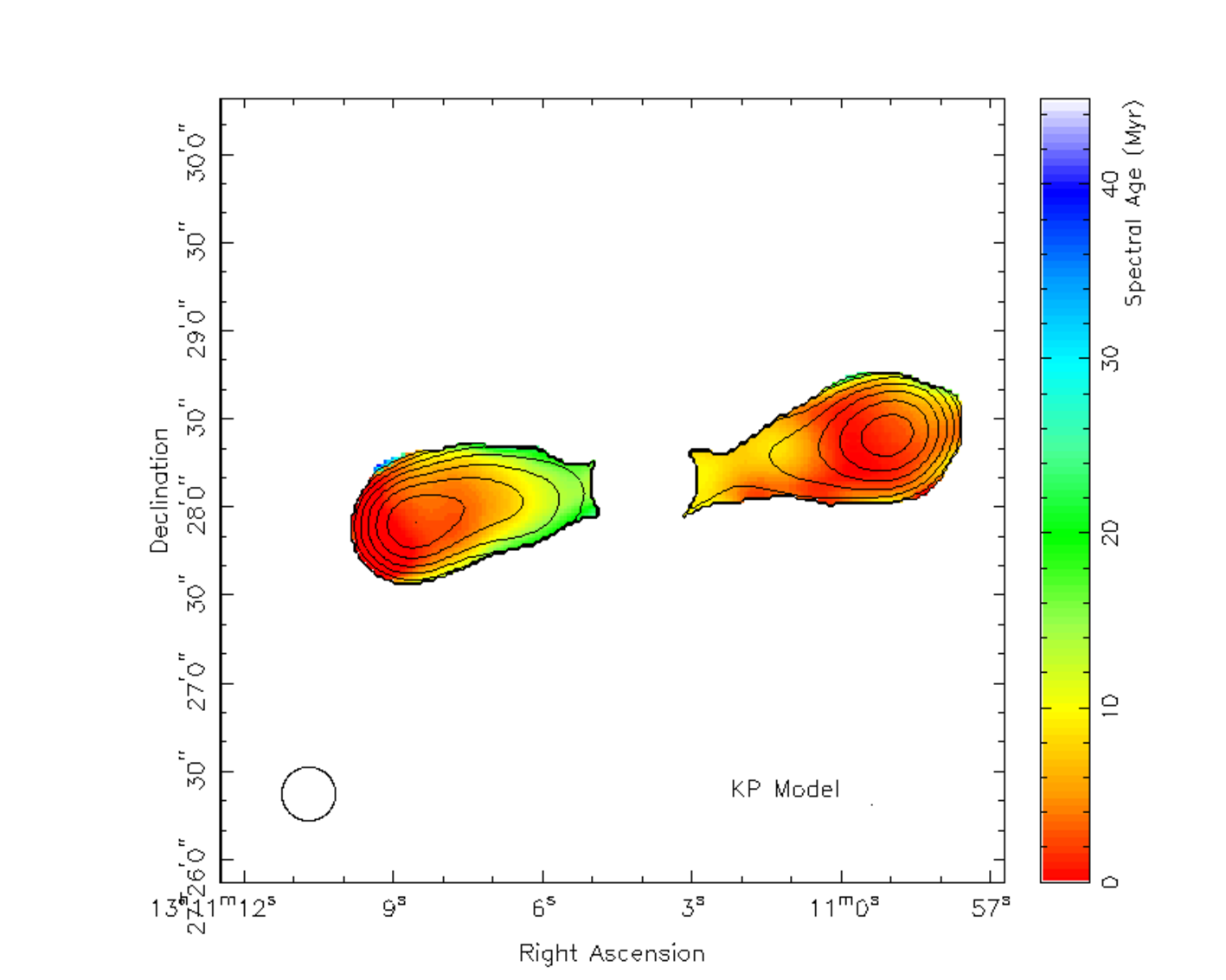}
\includegraphics[angle=0,width=8.8cm]{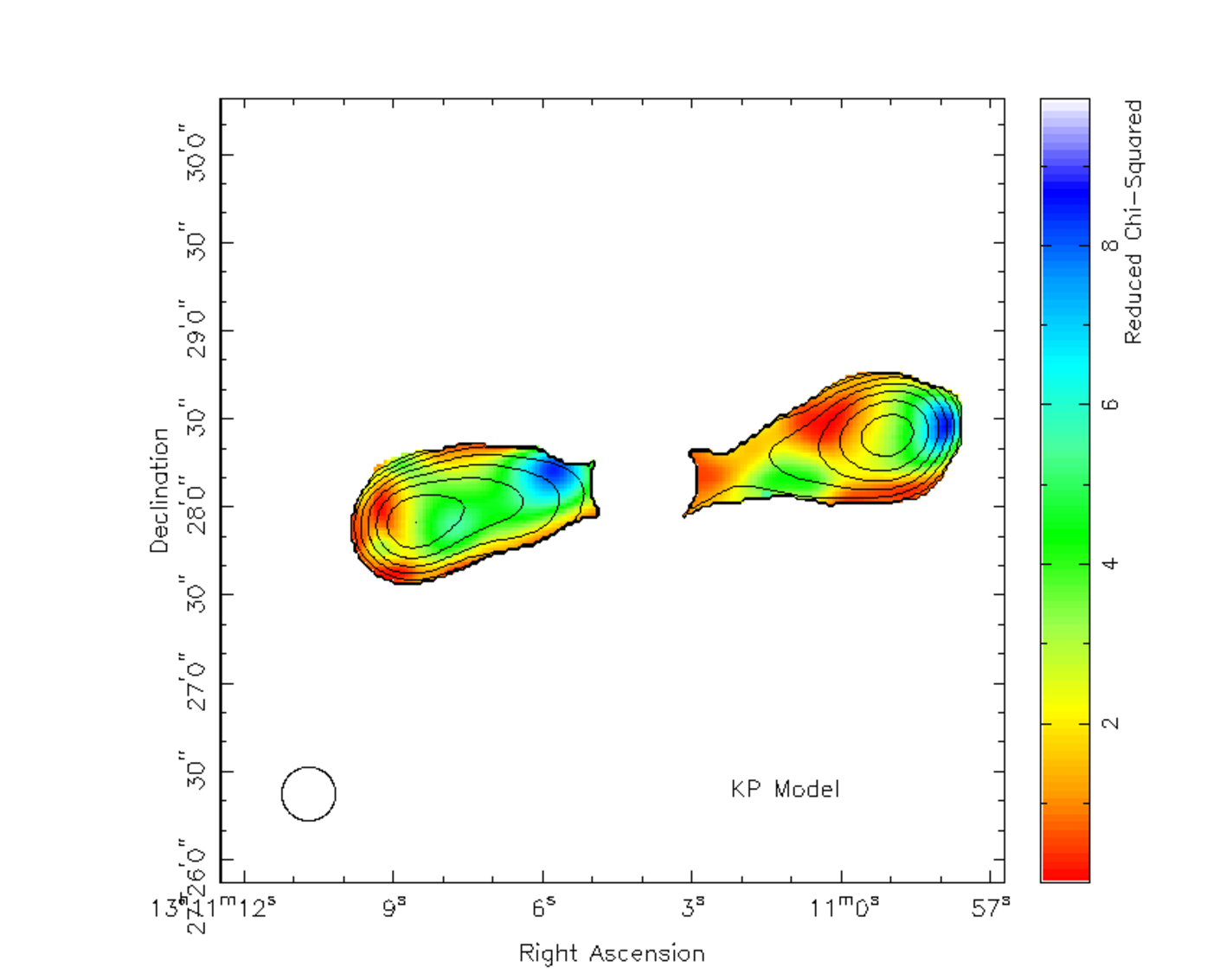}
\includegraphics[angle=0,width=8.8cm]{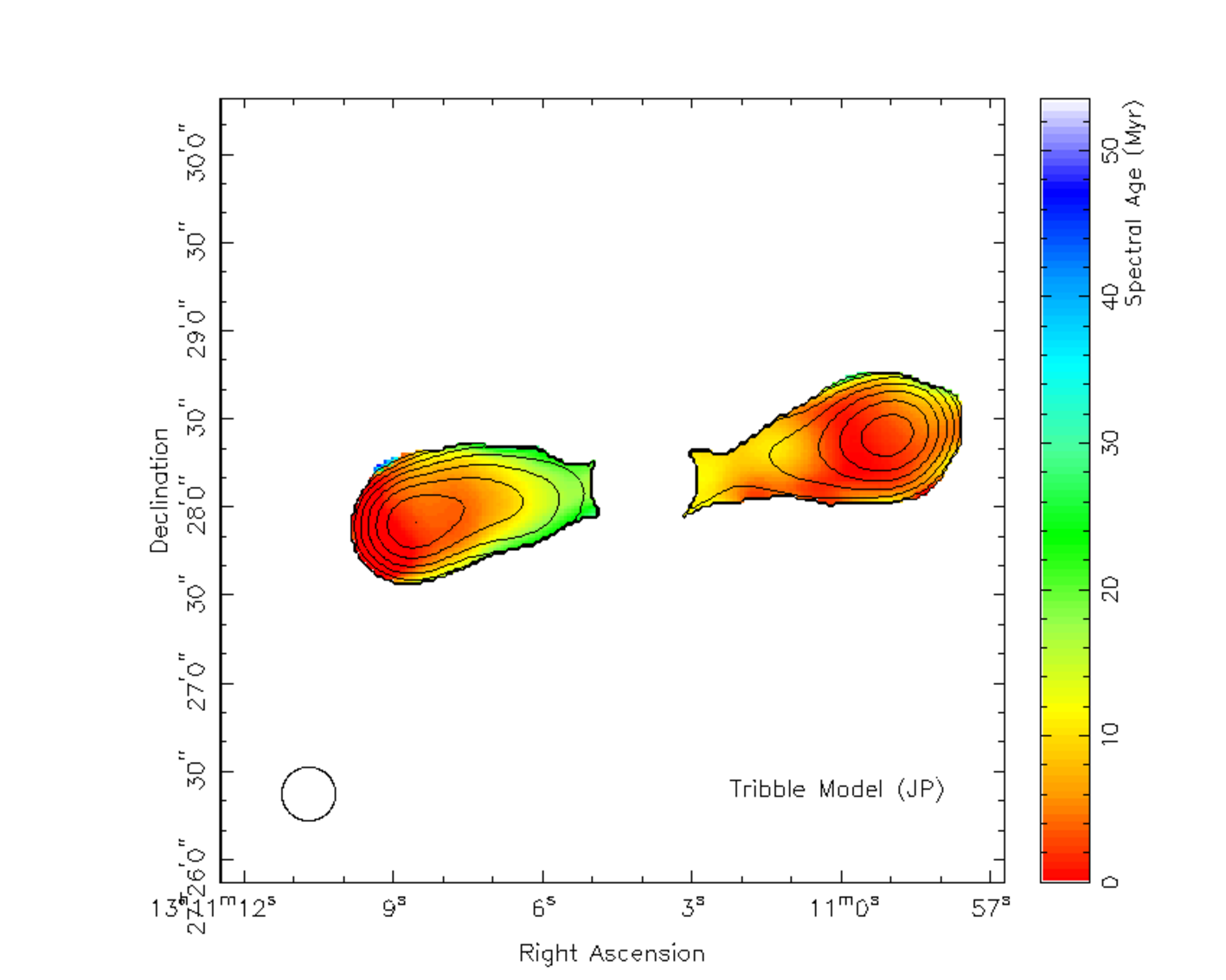}
\includegraphics[angle=0,width=8.8cm]{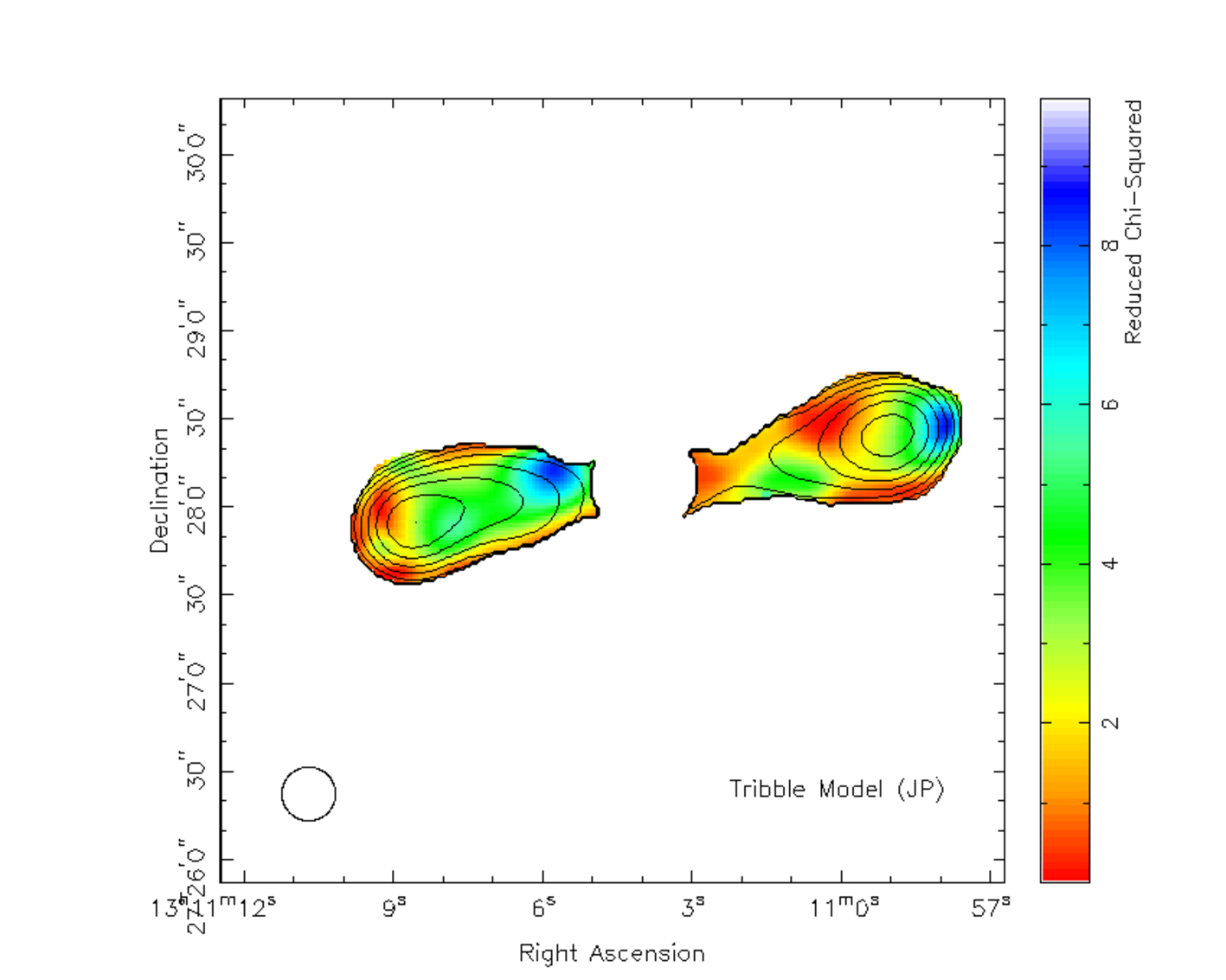}
\caption{Spectral ageing maps (left) and corresponding $\chi^2$ maps (right) of 3C 284 with 1.5 GHz flux contours. Three model fits are shown KP model (top), JP model (middle) and Tribble model (bottom) with an injection index 0.90.}
\label{spec-age-3C284}
\end{figure*}

\subsubsection{3C 284}
The spectral age and the reduced $\chi^2$ as a function of position for JP, KP and Tribble models are shown in Figure \ref{spec-age-3C284}. 
The spectral age of the hot-spots is lower compared to the rest of the lobes, and it is increasing towards the core. 
3C 284 also shows a classical behaviour in the nature of the age variation, where the lowest age regions are located close to the hot-spots, and there is a general trend of increasing while moving towards the galaxy core. The east direction jet is comparatively older than the west direction jet.

The reduced $\chi^2$ map as a function of position as obtained with the JP, KP, and Tribble models is shown in the right panel of Figure \ref{spec-age-3C284}.
It is observed that the $\chi^2$ varies within the range of 5, indicating a good sign of fitting. The $\chi^2$ exceeds 8 in the two regions of both the east and west sides.
The comparatively poor model fit in these areas is due to the edge effects and the dynamic range issue.

The KP, JP, and Tribble models of spectral ageing have been a longstanding basis for the spectral analysis of the lobes of FR-II radio galaxies. Therefore, it is not surprising that the best-fitting injection indices provide similar results for all model fits. The basic physical principles underlying the models are nearly the same. All three models assume that the radio emission from a source is generated by synchrotron radiation from relativistic electrons in a magnetic field. \citet{Ha13} also found a similar result for two sources using three different models. However, it is important to note that the accuracy of the age estimates depends on the quality and quantity of the observed data and the proper application of the models.

\subsection{Jet Power} 
In FR II radio galaxies, particles are accelerated by the jet-termination shock (JTS) at the lobes. 
The efficiency of the acceleration depends on the strength of the JTS, which is supposed to depend on jet power ($Q_j$).  
We have adopted the method from \citet{Ko13} for estimating the jet power. 
The jet power is calculated by estimating the total enthalpy of the lobes divided by the spectral age of the source.  
The expression for jet power can be written as $Q_j=\frac{4PV}{t_{active}}$ \citep{Ko13}, where $P$ and $t_{active}$ are the pressure in the lobe and time-scales of the active phase of the jets, respectively, assuming the minimum energy condition. For the value of $t_{active}$, the spectral age determined using JP models is used. We have used the calculated magnetic field ($B_{min}$) to estimate the overall pressure $P (= \frac{1}{3}(U_{B,min}+U_{e,min}))$ of the lobes, where minimum energy density in electrons be $U_{e,min}$ and that in the magnetic field be $U_{B,min}$.
For estimating the volume of the lobes, the cylindrical geometry of the lobe is used.
The jet powers ($Q_j$) for the galaxy 3C 35 for northern and southern lobes are estimated to be $7.6\times10^{44}$ erg sec$^{-1}$ and $8.1\times10^{44}$ erg sec$^{-1}$ respectively, while for 3C 284, the jet power for the eastern and western lobes is found to be $4.1\times10^{46}$ erg sec$^{-1}$ and $2.6\times10^{46}$ erg sec$^{-1}$ respectively.

\subsection{Lobe Speed} 
The advance speed of the lobes can be calculated using the standard procedure used by \cite{Al87}. For calculating the lobe speeds, it is assumed that the electron population accelerated from the hot-spot. The hot-spot is the area of intense emission at the end of the radio lobes. The advance speed of the lobe can be determined by $v_{\textrm{lobe}}=d_{hs}/t_{old}$, where $v_{\textrm{lobe}}$ is the lobe speed, $d_{hs}$ is the distance of the hot-spot, and $t_{old}$ is the age of the oldest region of the plasma.

In Table \ref{lobespeed}, the advance speed of the lobes is tabulated. These results are within the range expected from the earlier analysis by \citet{Al87}.
We find the advance speed for 3C 35 to be between 1.4 (Tribble) and 1.6 (JP) $\times 10^{-2}$ for the northern lobe and 1.5 (Tribble) and 1.7 (KP) $\times 10^{-2}$ for the southern lobe. The variation in the advance speed is not much between the northern and southern lobes. This small difference in the advance speed may be due to the effects of both orientation and the jet environment of the two lobes.

\begin{table}
\centering
\scriptsize
\caption{Lobe Advance Speeds}
\label{lobespeed}
\begin{tabular}{lccccc}
\hline
\hline
Source    & Model  & Lobe    & Max Age  & Distance  & Speed  \\
          &        &         & (Myrs)   & (kpc)     & ($10^{-2}$ $v/c$) \\
\hline
3C 35     & JP     & North   & 79.5     & 386.3    & 1.6 \\
          & KP     & North   & 80.9     & 386.3    & 1.5 \\
          & Tribble& North   & 86.9     & 386.3    & 1.4 \\
          & JP     & South   & 73.5     & 374.6    & 1.6 \\
          & KP     & South   & 72.0     & 374.6    & 1.7 \\
          & Tribble& South   & 79.5     & 374.6    & 1.5 \\
3C 284    & JP     & North   & 12.4     &  59.6    & 1.5 \\
          & KP     & North   & 18.3     &  59.6    & 1.0  \\
          & Tribble& North   & 16.8     &  59.6    & 1.1  \\
          & JP     & South   & 12.4     &  62.8    & 1.6 \\
          & KP     & South   & 18.3     &  62.8    & 1.1 \\
          & Tribble& South   & 16.8     &  62.8    & 1.2 \\
\hline
\end{tabular}
\end{table}

\section{Conclusion}
We have presented the multi-frequency radio observations of two large FR II radio galaxies, 3C 35 and 3C 284, using the low-frequency GMRT data and the high-frequency VLA data. We have studied the morphology of the galaxies at different frequencies. We searched for episodic jet activity in these two galaxies and found no indications of this kind of activity. Spectral ageing analysis of these two radio galaxies is performed. Two widely studied models for spectral ageing, namely, the KP and JP models, are used. The more realistic and complex Tribble model is also used. The spectral ageing map for each source is presented, where the age variation is shown as a function of position. We noticed that the age of the lobes of galaxy 3C 35 varies from 40 Myr to 60 Myr at different locations of the jet, while for galaxy 3C 284, the age of the lobes varies between 5 Myr and 20 Myr. The spectral age of hot-spots is lower compared with the other parts of the jet, and it is increasing along with the jet towards the core, which is a natural age variation for an FR II radio galaxy.
Different physical parameters of the two sources, like the magnetic field, jet power, and lobe speed, are estimated.
The estimated magnetic fields for 3C 35 and 3C 284 are $1.1 \times 10^{-9}$ Tesla and $3.0 \times 10^{-9}$ Tesla respectively.
The jet powers of 3C 35 for the northern and southern lobes are found to be $7.6\times10^{44}$ erg sec$^{-1}$ and $8.1\times10^{44}$ erg sec$^{-1}$ respectively, while for 3C 284, the jet powers for the eastern and western lobes are estimated to be $4.1\times10^{46}$ erg sec$^{-1}$ and $2.6\times10^{46}$ erg sec$^{-1}$ respectively.We determine the advance speed of the lobes using the standard method employed by \citet{Al87}.

\section*{Acknowledgement}
DP acknowledges the post-doctoral fellowship of the S.~N.~Bose National Centre for Basic Sciences, Kolkata, India, funded by the Department of Science and Technology (DST), India. We thank the anonymous reviewer for his/her helpful suggestions. We have used data from GMRT and VLA. The GMRT is a national facility operated by the National Centre for Radio Astrophysics of the Tata Institute of Fundamental Research. We thank the staff for helping with the observations. The VLA is run by the National Radio Astronomy Observatory (NRAO). NRAO is a facility of the National Science Foundation operated under a cooperative agreement by Associated Universities, Inc.  


\end{document}